\documentstyle[epsfig,12pt]{article}
\begin{document}
\title{Charmonium Spectrum from Quenched Lattice QCD with
       Tadpole Improvement Action}

\author{{Da Qing Liu}\\
        {\small Institute of Theoretical Physics, Chinese Academy
                of Sciences}\\
       }
\maketitle

\begin{center}
\begin{minipage}{5.5in}
\vskip 0.8in
{\bf Abstract}\\
We report here our lattice simulation on the charmonium spectra
in the quenched approximation.
Because the full adjustment on the nonperturbative parameters
such as $C_E$, $C_B$, $m_0a_s$ and $r_s$ needs
many calculation time, we only adjust two of them, $m_0a_s$ and
$r_s$( $\xi_3$ plays this role in the paper) but with some rescale
for mass splitting. After the rescale, we find that our results
are in agreement with the experiment ones.

\end{minipage}
\end{center}
\vskip 1in
\indent

\newpage
\section {Introduction}
Due to the importance of the charmonium spectra in particle
physics, there are many jobs on it in  lattice QCD. These
jobs did their ways with NRQCD approach$^{\cite{nrqcd}}$ or with
Wilson action with clover-improvement
$^{\cite{ping}\cite{jpn}\cite{ukqcd}\cite{taro}}$.

We report here our lattice results on the charmonium spectra
on anisotropic lattice ( we will also present our results on the
bottomonium spectra in the future). Paralleling with references
\cite{ping}-\cite{taro}, our results were obtained with the
Sheikholeslami-Wohlert action$^{\cite{swa}\cite{lm1}}$(SW action)
as well as with the tadpole improvement. As many authors pointed
out, such improvements make us get interesting results with the
fewer calculation. In fact, our simulation was done on personal
computations.

As pointed out in reference \cite{kronfeld}, if we adjust
parameters to take the right hyperfine splitting in this action,
the mass splitting generated by the orbit excitation, including
orbit angular momentum and therefore spin orbit interaction,
will be incorrect (we call this mass splitting s.o. splitting
thereinafter). Therefore, to get the correct
charmonium spectra, or the mass
splitting, some argument on the mass dependent choice of $C_E$
was done in the article \cite{kronfeld}. In fact, the choice of
the clover coefficient have been performed in various ways, which
include applying different orders in perturbation
theory$^{\cite{per}}$
and introducing non-perturbative improvement criteria using
Schr$\ddot{o}$dinger function$^{\cite{nonp}}$. Those schemes,
however, need much more computer or calculation time.

If parameters on lattice are not maladjusted so much,
then the mass splitting and the gross mass, which is mainly
dictated by the static quark mass $M_1$, are in different scale.
To make the scale at the same, maybe some
rescale scheme for the measured mass splitting is also proper.
After the rescale, we find that our results are in good
agreement with the experiment ones.

The section 2 shows some notations we should considered when
we begin to calculate, and the section 3 and 4 is our simulation detail
and results.  At last, we present the summary in the section 5.

\section {The Choice of Parameters for Simulation}

To perform the simulation, we adopt the tadpole improvement
action for gluon as follows:
\begin{equation} \label{e1}
  S_G=\beta \{ \frac{5}{3} \frac{P_{sp}}{\xi u_s^4} +\frac{4}{3}
  \frac{\xi P_{tp}}{u_s^2}-\frac{1}{12}\frac{P_{sr}}{\xi u_s^6}
  -\frac{1}{12}\frac{\xi P_{str}}{u_s^4} \}
\end{equation}
where $P_{sp}$($P_{st}$) is the sum of all spatial(temporal)
plaquettes, $P_{sr}$ represents $2\times 1$ spatial rectangular
Wilson loops and $P_{str}$ is the Wilson loops with 2 spatial
and one temporal links. $\xi$ is the aspect ratio:
\begin{equation}\label{e2}
\xi =\frac{a_s}{a_t}.
\end{equation}
This gluonic action, as many authors pointed out$^{\cite{ga}}$, may improve
the action up to $O(a^4_s,a^2_t)$.

For quarks, we adopt the SW action.
This action was proposed initially to improve the continuum behavior
for the fermions. But as suggested by l$\ddot{u}$scher and
Weisz$^{\cite{luscher1}}$, one had better take the scheme with
the on-shell improvement. This scheme had been illuminated
systematically in reference \cite{kronfeld} in action form as well as in
Hamilton form. Here we shall use the scheme to determine some
parameters in SW action.

We take the SW action as:
\begin{eqnarray}
S_q & =& \sum\limits_x \bar{\psi}(x)\{
m_0 a_s+\frac{1}{2}[ \xi_1 \gamma_4 \nabla_4 +\sum\limits_i \gamma_i
\nabla_i (1-\frac{1}{6}\triangle_i)-\xi_1 \triangle_4
\nonumber \\
~~& ~~& -\frac{1}{\xi_3}\sum\limits_i(\triangle_{i}-\frac{1}{12}
\triangle^2_i)+C_E \sum\limits_i \sigma_{4i}\hat{F}_{4i}+C_B \frac{1}{\xi_3}
\sum\limits_{i>j} \sigma_{ij}\hat{F}_{ij}
] \} \psi(x),
\nonumber \\
\nabla_{\mu}\psi (x) &=& u_{\mu}(x)\psi(x+\hat{\mu})-u^{\dag}_{\mu}
(x-\hat{\mu})\psi (x-\hat{\mu}),
\nonumber \\  \label{e3}
\triangle_{\mu}\psi (x) &=& u_{\mu}(x)\psi(x+\hat{\mu})+
u^{\dag}_{\mu}(x-\hat{\mu})\psi(x-\hat{\mu})-2\psi(x),
\end{eqnarray}
where $m_0$ is the bare mass of the quark. We shall
determine  the
 parameters $\xi_1$, $\xi_3$, $C_E$ and $C_B$ in Eq.
 (\ref{e3}) in the following.
\subsection{The choice of parameters $\xi_1$ and $\xi_3$}
 The propagator for quark is
 \begin{equation}
 <\psi(t^{\prime},p^{\prime})
 \bar{\psi}(t,p)>=(2\pi)^3\delta(p^{\prime}-p)C(t^{\prime}-t,p),
 \end{equation}
where
\begin{equation}   \label{e4}
C(t,p)={\mathcal{Z}}(p) \frac{e^{-E |t|}}{\sinh~E}
\end{equation}
with the applying of the residue theorem. In Eq. \ref{e4},
${\mathcal{Z}}(p)$ is the function which is independent of time $t$.
The energy $E$ of the quark with momentum $p$ is determined by
\begin{eqnarray}    \label{e5}
G^2+w^2+\xi_1 &=& 2 G \xi_1 \cosh E a_t, \nonumber \\
G &=& m_0+\xi_1+\frac{1}{\xi_3} \sum\limits_i
(1-\cos p_i+\frac{1}{6}(1-\cos p_i)^2), \nonumber \\
w^2 &=& \frac{1}{9} \sum\limits_i (\sin p_i ~(4-\cos p_i))^2.
\end{eqnarray}
To get above equation, we set $a_s=1$. So, one may expand $G$
and $w^2$ according to the power of $p_i$:
\begin{eqnarray}
G &=& m_0+\xi_1+\sum\limits_i \frac{p^2_i}{2 \xi_3}+O(p^6),
\nonumber \\
w^2 &=& \sum\limits_i p^2_i +O(p^6).
\end{eqnarray}
At $m_0=0$, one gets:
\begin{equation} \label{e6}
(\xi_1 E a_t)^2+O(a^4_t)=\sum\limits_i p^2_i+O(p^4_i),
\end{equation}
where $a_t=1/\xi$.
Because of the Eq. (\ref{e6}), we choose $\xi_1=\xi$ with the tuning of the
parameter $\xi_3$ to make the following dispersion relation of the particle
$\psi$ be satisfied:
\begin{equation}             \label{e7}
E^2(J/\psi)=m^2(J/\psi)+c^2\sum\limits_i p^2_i,
\end{equation}
where $c=1$ is the effective velocity of photon(EVP).

One can not make all the dispersion relations of the calculated particle
be satisfied at the same time with the tuning of the sole
parameter. Phenomenally, this infraction manifests the fact that the
relative movement of two constituent quarks in meson is various with
different meson( See appendix for more detail).

\subsection{The Choice of $C_E$ and $C_B$}
The key points of the choice for the clover coefficient $C_B$ is
the isospectrum transformation and the redundant coupling.
After some algebra calculation, one can easily see that the choice
$C_B=1$ is proper$^{\cite{kronfeld}}$.

There are a few choices of $C_E$. For instance, one may choose
$C_E=1$ or $C_E=0$ $^{\cite{lepage}}$ in NRQCD.
In fact, following the method developed in reference
\cite{kronfeld} and \cite{jpn}, one may see that
in this action the parameters should satisfy
\begin{eqnarray}      \label{on-tree}
C_B &=& 1,     \nonumber \\
\xi_3 &=&
\frac{\xi}{1+m_t}(\frac{1}{\ln(1+m_t)}-\frac{2}{m_t(2+m_t)})^{-1},
\nonumber \\
C_E &=&
\frac{\xi}{\xi_3(1+m_t)}+(\frac{\xi}{\xi_3(1+m_t)})^2\frac{m_t(2+m_t)}{4},
\end{eqnarray}
where $m_t=\frac{m_0a_s}{\xi}$. The above condition is at tree
level but with tadpole improvement. While for getting the
nonperturbative parameters we should adjust these parameters with
the same physical principle that led to Eq. (\ref{on-tree}). For instance,
to get the nonperturbative $\xi_3$ we adjust $\xi_3$ to make
Eq. (\ref{e7}) to be satisfied.

We select $C_B=1$ for our simulation. While
noticing that $C_E=\xi/\xi_3$
at vanished $m_t$ we just take a mass
independent setting for $C_E$ for simplify:
\begin{equation}   \label{xi3}
C_E=\frac{\xi}{\xi_3}.
\end{equation}
The choice in Eq. (\ref{on-tree}) and (\ref{xi3}), however,
is at tree level and sometimes a little maladjusted.
To compensate the maladjustment
will lead to a mass splitting rescale.

Thus we have only two parameters $\xi_3$ and $m_0$ to be
determined in the simulation: one is adjusted to give the correct
energy-momentum relation for $J/\psi$ and the other will be
rectified to determine the lattice spacing, $a_s$($a_t$).

\section{The Detail for the simulation}
In this work, we calculate five meson masses of charmonium, named
as $^1S_0(\eta_c)$, $^3S_1(J/\psi)$, $^1P_1(h_c)$,
$^3P_0(\chi_{c0})$ and $^3P_1(\chi_{c1})$ and their excited states.
The choice of the operator for definite meson is as the standard
one$^{\cite{ping}}$. These operators are called $\Gamma$
ones in ref. \cite{jpn}.

To determine $\xi_3$, we calculate the energy for four low-lying momentum
$\frac{L}{2\pi}{\mathbf{p}}a_s=(0,0,0)$, $(0,0,1)$, $(0,1,1)$ and
$(1,1,1)$ of the particle $J/\psi$ and fit them by
\begin{equation}\label{e8}
(E(|{\mathbf{p}}|)a_t)=(E(0)a_t)^2+\frac{c^2_0}{\xi^2}(|{\mathbf{p}}|a_s)^2.
\end{equation}
We tune $\xi_3$ to make the relation $c_0=1$ be satisfied in one
percent accuracy.

There are types of scheme to determine $m_0$,or equivalent $a_t$.
For instance, one may determine $m_0$ based on the hyperfine splitting
or based on the s.o. splitting. As for the  first type,
One may take the ratio of the masses of S-state mesons:
\begin{equation} \label{e9}
\frac{m(\psi^{\prime}) m(\eta_c)}{m(\eta_c^{\prime})m(J/\psi)}=0.987.
\end{equation}
And as for the second, one may take, for example,
the ratio of the mass of $\chi_{c1}$ and $J/\psi$
\begin{equation} \label{e91}
\frac{m(\chi_{c1})}{m(J/\psi)}=1.134.
\end{equation}

Then, we can extract $a_t$ by
\begin{equation}       \label{e10}
a_t=\frac{m({1\bar{S}})a_t}{m({1\bar{S}})_{exp}},
\end{equation}
where
$m({1\bar{S}})=(m(1^1S)+3m(1^3S))/4$ is the measured average
mass of $1^1S$ and $1^3S$ state on the lattice and $m(1\bar{S})_{exp}$
is the experimental one, $3.0676Gev$.

In this paper we iterate $m_0$ and $\xi_3$ to make both the
equation (\ref{e8}) and (\ref{e9}) be hold at the same time.

Because of our computational limitation, we did our simulation on a
$L\times T=6^3 \times 36$ lattice with $\xi=6.0$.
Parameters for the simulation are displayed in table 1.

For a definite $\Gamma$ operator we measure its correlation function as:
\begin{equation}
C_{state}(t,{\mathbf{p}})=\sum\limits_{{\mathbf{x}}}<\bar{\psi}
({\mathbf{x}},t)
\Gamma \psi({\mathbf{x}},t)\bar{\psi}({\mathbf{0}},0)\Gamma
\psi({\mathbf{0}},0)>e^{-i \mathbf{p\cdot x}}.
\end{equation}
The energy can be extracted by n-hyperboliccosine  ansatz( n is the number
of the extracted masses and T is the periods in the time direction):
\begin{equation}
C_{state}(t)=\sum\limits^{n-1}_{j=0}a_j(e^{-m_jt}+e^{-m_j(T-t)}),
\end{equation}
which means that we should find parameters $a(j)$ and $m(j)$ to minimize
the function
\begin{equation}                \label{e13}
\sum\limits_{i=l_i}^{l_f} W_i (C(t_i)-\sum\limits^{n-1}_{j=0}
a_j(e^{-m_jt_i}+e^{-m_j(T-t_i)}))^2/C(t_i)^2,
\end{equation}
where $W_i\propto \frac{C(t_i)^2}{{\triangle_{C(t_i)}}^2}$ is the
weight of the correlation $C(t_i)$.

We choose $n=2$ for all the measured particles in the work.
Apparently, the excited states we measured here is not the real
first excited states in nature but the mixture, the mainly
component of which is the first and the second excited states.
Thus all the measured excited masses are larger than the
experiment ones. We expect that the scheme we adopted in Eq.
(\ref{e91}) would decrease this mixing  impact.

For each $l_i$ and $l_f$ in Eq. (\ref{e13}) one may find a
group of corresponding parameters $m_j$ and $a_j$. Then, one may choose
a set of parameters $l_i$, $l_f$, $m_j$ and $a_j$, which make the
function $\sum\limits_j
(\frac{\triangle^2_{m_j}}{m_j^2}+\frac{\triangle^2_{a_j}}{a_j^2})$
get its minimal points for all the sets. But due to the fact that
nearly all the minimal point for a definite operator is
approached at the set in which $l_i=1$ and $l_f=T-1$, we fix our
set in which $l_i=1$ and $l_f=T-1$. This is as expected, for the
more  measured data means the more information if we excluded
the case of $l_i=0$ or $l_f=T$, in which the counterterms will
enter in.

\section{Simulation Results}
In table 2, we list the results of the typical dispersion relation
and the fitting data for low-lying particles  at $\beta=3.0$ and
$m_0a_s=1.06$.

From the simulation we know that the EVP
for the particle $\eta_c$ is equal to the one of the particle
$J/\psi$ in about 2 percent. This is obvious from the
appendix since their mass splitting is mainly due to the spin
splitting$^{\cite{stephen}}$. The similar result is obtained by
other authors$^{\cite{ping}-\cite{ukqcd}}$.

On the other hand, we find that the EVP in
different mesons usually have the relation:
\begin{equation}
c(h_c)\leq c(\chi_{c1}) \leq c(\chi_{c0}) \leq c(J/\psi) \approx
c(\eta_c).
\end{equation}
This possibly implies that the expanding point $p_0$s, or roughly saying,
the main distribution of the relative momentum, can be estimated
as
\begin{equation}
p_0(h_c)\geq p_0(\chi_{c1}) \geq p_0(\chi_{c0}) \geq
p_0(\psi) \approx p_0(\eta_c).
\end{equation}

From the data we see that the EVPs for all
particles except meson $\eta_c$ and $J/\psi$ are usually smaller than unity,
which means that the kinetic mass $M_2$
is larger than the static mass $M_1$( see detail in ref. \cite{kronfeld}).
At the same time, to get the correct $\chi_J$ mass splitting, ref.
\cite{kronfeld} suggested a mass dependence of $C_E$, which leads
to a scheme of adjustment of three parameters: $m_0a_s$, $\xi_3$ and
$C_E$.

To combine the scale of $M_1$, $M_2$ and $M_E$ and decrease
the number of adjusted parameters, we adopt such progress:
we adjust only two parameters  $m_0a_s$ and $\xi_3$
but with the mass splitting rescale.

We take Eq. (\ref{e9}) to determine $m_0a_s$ and measure the mass of
particle $\eta_c$ and $J/\psi$, but for other particle, since we
should consider the s.o. splitting, we redefine its mass as follows.

We rescale the mass splitting between $1^3P_1$ and averaged $1S$
state as
\begin{equation}       \label{e15}
m(\chi_{c1})-{m(1\bar{S})}=0.443Gev .
\end{equation}
Therefore, the modified mass for a definite meson is
\begin{equation} \label{e16}
m_{mod}=0.443\frac{ma_t-m(1\bar{S})a_t}
{m(\chi_{c1})a_t-m(1\bar{S})a_t}+m(1\bar{S})~(Gev),
\end{equation}
where $ma_t$ is the dimensionless measured mass on the lattice and
$a_t$ is extracted by Eq. (\ref{e10}).

The simulation parameters are listed in table 3.
Our results for scale are not inconsistent with the result
obtained from the Sommer scale $r_0$ in the calculation of the
gluon spectra$^{\cite{morningstar}}$ although they are
achieved in very different aspect.

We list our modified charmonium spectra in table 4, which is,
as well as in other references, in agreement with the experiment.
From the data
we find that at the finite lattice spacing the ratio
$\frac{m(J/\psi)}{m(\eta_c)}$ is smaller than the experiment one.
To get the correct ratio one may decrease $m_0$ and therefore achieved
a decreased lattice spacing, $a_s$. An estimation for the
decreased $m_0$ shows that it approximately leads to the same
result as in reference \cite{morningstar}. To decrease the
computer time, we take the scale determined by Eq. (\ref{e9}) with
the notice that the ratio between $a_s$ obtained by us and that
obtained by C.J. Morningstar is nearly the same at different $\beta$,
i.e., the ratio is 1.6 at $\beta=3.0$ and 1.7 at $\beta=2.5$.

While compared with our results with those in \cite{taro}, one may
find, that our result for hyperfine splitting is smaller than that
in \cite{taro}. Maybe the smallness is caused by the too large
lattice spacing and/or the contamination of hyperfine splitting
by the s.o. terms in the determination of $m_0a_s$ in
Eq. (\ref{e9}).

For the known excited states, our results,
as those  obtained in ref. \cite{jpn}, are larger than the experiment ones.
But for unknown states, it needs some more explore to explain
why our result are also larger than those in ref. \cite{jpn}.

While following the reference \cite{kronfeld}, one should do this
rescale between $\chi_J$ states. But, from the results we find that
our rescale scheme also obtains the correct $\chi_J$ states.

\section{Conclusion}
For the simulation of the charmonium spectra, because the ratio
between the rest mass $M_1$ and the kinetic $M_2$ are not always
the same, so the scale for mass splitting $\Delta m$ and the the
gross mass $m$ is different. Since $\Delta m$ is not mainly dictated
by the static mass $M_1$, reference \cite{kronfeld} suggested the
following strategy: forget about $M_1$ and adjust the bare mass so
that the kinetic mass $M_2$ takes the physical value.

To make mass splitting and the gross mass at the same scale,
one may adjust parameters or rescale mass splitting.
This rescale combines the scale of
the gross $M$ and the mass splitting, $\Delta M$.
After the rescaling, we get the results which are in
agreement with the experiment ones despite a little
maladjustment of the parameters.

\newpage
\section*{Appendix A}
\setcounter{equation}{0}
\renewcommand{\theequation}{\roman{equation}}
 From phenomena, one may regard the meson consist of one couple
 of constituent quark and antiquark and one can also in principle
 write the wave function for (anti-)charm in momentum and spin space.
The relative movement between them depicts the different mesons.
For instance, the orbital angular momentum in $\eta_c$ is S-wave
while that in $h_c$ is P-wave. On the other hand, for $c\bar{c}$
system the typical velocity in particle is $v^2\approx 0.3$, i.e.
the energy scales
 $m_c v^2/2$ is about 200-800 Mev. This implies that the mass
 splitting generated by the relative movement between quark and
 antiquark is mainly dictated by the dynamical behaviour of quarks
 at the vicinity around $m_cv^2/2$. So, maybe the better
 expanding points in meson is $p^2_0(/2m_c)$ rather than $p^2_0=0$,
 where $p^2_0$ is phenomenal parameter to describe the relative
  movement of quarks in mesnon.

 In the following we only consider the effect of the relative
 movement in 1+1 dimension lattice.

 In continuum, the energy-momentum relation for the quark
 is:
 \begin{equation}
 E^2(p)=m^2+\frac{c^2 p^2}{\xi^2}=m^2+\frac{c^2}{\xi^2}(p_0^2+\triangle p^2),
 \end{equation}
where all the quantity is dimensionless one, i.e. $E$ is $E a_t$,
quark mass $m$ is $m a_t$ and $p$ is $p a_s$.
One may rewrite it as:
\begin{equation} \label{a1}
e^E+e^{-E}=e^{m \alpha}+e^{-m\alpha}+\frac{c^2 \triangle p^2}{2
m\alpha \xi^2}(e^{m\alpha}-e^{-m\alpha})+O((\triangle p^2)^2),
\end{equation}
where $ \alpha ^2=1+\frac{c^2 p^2_0}{m^2 \xi^2}=1+\gamma^2 v^2_c$.
One may redefine the phenomenal parameter $v^2$ to set $\gamma=1$.

On the other hand, on an 1+1 dimension lattice we have:
\begin{eqnarray}
w^2 &=& w^2_0+f_1 \triangle p^2,
\nonumber \\
w^2_0 &=& \frac{1}{9}\sin ^2p_0 (4-\cos p_0)^2,
\nonumber \\
f_1 &=& \frac{\sin p_0 (4-\cos p_0)}{9 p_0}(4\cos p_0-\cos 2p_0);
\end{eqnarray}
and
\begin{eqnarray}
G &=& G_0+f_2 \triangle p^2+O((p^2)^2),
\nonumber \\
G_0 &=& m_0+\xi+\frac{2}{\xi_3}\sin^2p_0/2 (1+\frac{1}{3}\sin^2
p_0/2),
\nonumber \\
f_2 &=& \frac{\sin p_0}{6 \xi_3 p_0}(5-2\cos p_0).
\end{eqnarray}
Therefore, according to Eq. (\ref{e5}), on lattice one gets
\begin{equation}
e^E+e^{-E}=\frac{G_0}{\xi}+\frac{\xi}{G_0}+\frac{w^2_0}{\xi G_0}
+(\frac{f_2}{\xi}+\frac{f_1}{\xi G_0}-\frac{f_2}{G^2_0}
(\xi+\frac{w^2_0}{\xi}))\triangle p^2+O((\triangle p^2)^2).
\end{equation}
Comparing with Eq. (\ref{a1})and setting $c=1$, one knows that
\begin{eqnarray}       \label{a3}
\xi_3 &=& \frac{1-t^{-2}(1+w^2_0/\xi^2)}{\frac{\sinh m\alpha}{m\alpha}
-\frac{f_1}{t}}f_2 \xi_3 \xi,
\nonumber \\
m_0 a_s&=& \xi(t-1-\frac{2}{\xi \xi_3}g),
\end{eqnarray}
where
\begin{eqnarray}
g &=& \sin^2 p_0/2 (1+\frac{1}{3}\sin^2 p_0/2),
\nonumber \\
t &=& \frac{b+\sqrt{b^2-4(1+w^2_0/\xi^2)}}{2},
\nonumber \\
b &=& e^{m\alpha}+e^{-m\alpha} .
\end{eqnarray}
While noticing that functions $f_2$, $f_1$, $G_0$ and $w^2_0$ are
even ones for $p_0$, one may coarsely replace $p_0$ with $\sqrt{p^2_0}$
in those functions.
We expand Eq. (\ref{a3}) with respect to $ma_s$:
\begin{eqnarray} \label{a2}
\xi_3 &\approx& (1-2\frac{1+\gamma^2v_c^2}{3}\frac{m a_s}{\xi})\xi,
\nonumber \\
m_0 a_s &=& ma_s+O((ma_s)^3)
\end{eqnarray}
To applying the above equations, we should consider the intricate
additive and multiplicative renormalization$^{\cite{luscher2}}$.

The above argument is somewhat pedagogical, but it manifests the
correct qualitative behavior, even on the 3+1 dimension lattice(see
detail in the paper). For instance, from Eq. (\ref{a2}) it
is easily understood that the error of the pole mass will be proportion to
$ma$ if we choose $\xi_3=\xi$ naively.

\newpage
\vskip 0.5in
\begin{center}
\begin{tabular}{|cccc|}
\hline
$ \beta$ & $u^4_s$ & sweep/confg. & confgs. \\ \hline
2.5   & 0.423  & 40  & 100  \\ \hline
2.8   & 0.463  & 40  & 100  \\ \hline
3.0   & 0.486  & 40  & 100  \\ \hline
\end{tabular} \\
{\small {\bf Table. 1}~ Parameters for the simulation with the scale
set by Eq. (\ref{e9}).}
\end{center}

\vskip 0.5in
\begin{center}
\begin{tabular}{|ccccc|}
\hline
$(\frac{L}{2\pi}|{\mathbf{p}}|a_s)^2$ & 0 & 1 & 2. & 3 \\ \hline
$\eta_c$    & 0.276(3)  & 0.309(2)  & 0.340(2)  & 0.369(2)  \\
~           & 0.278     & 0.308     & 0.339     & 0.370  \\ \hline
$J/\psi$    & 0.286(1)  & 0.317(1)  & 0.349(1)  & 0.3775(13)\\
~           & 0.286     & 0.317     & 0.348     & 0.3786  \\ \hline
$h_c$       & 0.409(13) & 0.400(10) & 0.416(8)  & 0.432(7)  \\
~           & 0.398     & 0.408     & 0.419     & 0.429  \\ \hline
$\chi_{c1}$ & 0.401(9)  & 0.414(8)  & 0.432(7)  & 0.448(7) \\
~           & 0.400     & 0.416     & 0.432     & 0.448  \\ \hline
$\chi_{c0}$ & 0.380(11) & 0.405(11) & 0.428(12) & 0.450(12)\\
~           & 0.380     & 0.404     & 0.427     & 0.451 \\
\hline \hline
\end{tabular} \\
{\small {\bf Table. 2}~
$(E a_t)^2$ v.s. $(\frac{L{\mathbf{p}}a_s}{2\pi})^2$ for different
mesons. The data in upper row are measured ones and the data in
lower row are fitted ones. The fitting parameters $a$ and $b$
of the formula $(Ea_t)^2=a+b\frac{({\mathbf{p}}a_s)^2}{\xi^2}$
for different meson are 0.2776(20) and 1.007(34) for $\eta_c$,
0.2863(12) and 1.010(17) for $J/\psi$, 0.398(7) and 0.34(11) for
$h_c$, 0.3995(58) and 0.530(93) for $\chi_{c1}$, and 0.380(8) and
0.77(13) for $\chi_{c0}$ respectively. The corresponding
velocities of photon for different meson are: 1.003(17) for
$\eta_c$, 1.005(8) for $J/\psi$, $0.58(10)$ for $h_c$, $0.73(6)$ for
$\chi_{c1}$ and 0.88(8) for $\chi_{c0}$ respectively.
 }
\end{center}

\vskip 0.5in
\begin{center}
\begin{tabular}{|ccccccc|}
\hline
$ \beta$ & $\xi_3$ & $m_0a_s$ & $a_s(fm)$ & $La_s(fm)$ &
EVP($J/\psi$) & EVP($\eta_c$)   \\ \hline
2.5  & 3.31 & 2.33 & 0.322 & 1.93 & 1.000(10)& 1.022(21) \\ \hline
2.8  & 3.92 & 1.57 & 0.258 & 1.55 & 1.006(7) & 1.018(15) \\ \hline
3.0  & 6.38 & 1.06 & 0.208 & 1.25 & 1.005(8) & 1.003(17)  \\ \hline
\end{tabular} \\
{\small {\bf Table. 3}~ Parameters for the simulation with the scale
set by Eq. (\ref{e9}).
 }
\end{center}
\vskip 0.5in

\begin{center}
\begin{tabular}{|cccccc|}
\hline \hline
state & $\beta=2.5$  & $\beta=2.8$ & $\beta=3.0$ & $a_s\rightarrow 0$
& Exp.(Gev) \\ \hline
$\eta_c$             &3.030(11)  &3.032(11)  & 3.030(14) & 3.030(22) & 2.980 \\ \hline
$J/\psi$             &3.080(47)  &3.0795(44) & 3.080(7)  & 3.080(10) &3.097 \\ \hline
$h_c$                &3.553(135) &3.500(68)  & 3.539(43) & 3.546(117)& 3.526 \\ \hline
$\chi_{c1}$          &3.511(95)  &3.511(50)  & 3.511(32) & -         & 3.511 \\ \hline
$\chi_{c0}$          &3.471(139) &3.447(73)  & 3.436(46) & 3.412(119)& 3.415
\\ \hline \hline
$\eta_c^{\prime}$    & 4.309(26) & 4.347(17) & 3.982(15) & 3.766(31) & 3.594 \\ \hline
$\psi^{\prime}$      & 4.316(11) & 4.359(7)  & 3.993(6)  & 3.795(13) & 3.686  \\ \hline
$h_c^{\prime}$       & 4.748(107)& 4.964(68) & 4.633(58) & 4.564(126)& -     \\ \hline
$\chi_{c1}^{\prime}$ & 4.889(92) & 5.079(57) & 4.689(45) & 4.534(102)& -     \\ \hline
$\chi_{c0}^{\prime}$ & 4.980(150)& 5.145(94) & 4.721(70) & 4.510(161)& -
\\ \hline \hline
\end{tabular}    \\
{\small {\bf Table 4}~Results of the charmonium mass $M$ and the mass splitting
$\Delta M$\\ in units of Gev at $\xi=6.0$ with the same scale setting as in Table 1.}
\end{center}

\end{document}